\documentclass{revtex4}

\usepackage{graphicx}
\begin{document}
% You should use BibTeX and revtex.bst for references
\bibliographystyle{revtex}

% number  and your conference paper identification number on the
% title page in preprint mode. Multiple \preprint commands are allowed.
%\preprint{}

\title{Sensitivity to Doubly Charged Higgs Bosons in the Process 
$e^-\gamma\to e^+ \mu^-\mu^-$  }
% Optional argument for running titles on pages
%\title[]{}



\author{Stephen Godfrey}
\email{godfrey@physics.carleton.ca}
\author{Pat Kalyniak}
\email{kalyniak@physics.carleton.ca}
\author{Nikolai Romanenko}
\email{nromanen@physics.carleton.ca}
%\thanks{}
\affiliation{Ottawa-Carleton Institute for Physics \\
Department of Physics, Carleton University, Ottawa, Canada K1S 5B6}


\date{\today}

\begin{abstract}
We study the sensitivity to doubly charged Higgs bosons, 
$\Delta^{--}$, in 
the process $e^-\gamma\to e^+ \mu^-\mu^-$ for centre of mass energies 
appropriate to future high energy $e^+e^-$ 
collider proposals.
For $M_\Delta < \sqrt{s_{e\gamma}}$ discovery is likely for even
relatively  small values of the Yukawa coupling to leptons.  However, 
even far above threshold, 
evidence for the $\Delta$ can be seen due to contributions from 
virtual intermediate $\Delta$'s although, in this case,
$\mu^-\mu^-$ 
final states can only be produced in sufficient numbers for discovery 
for relatively large values of the Yukawa couplings.
\end{abstract}
% \pacs{PACS numbers: 12.15.Ji, 12.15.-y, 12.60.Cn, 14.80.Cp}

%\maketitle must follow title, authors, abstract and \pacs
\maketitle

\section{Introduction}
%\label{}

Doubly charged Higgs bosons would have a distinct experimental
signature. Such particles arise in many extensions of 
the Standard Model (SM) including as components of $SU(2)_L$ Higgs
triplets. Models with triplet representations include the Left-
handed Higgs triplet model of Gelmini and Roncadelli \cite{Gelmini},
where they provide Majorana masses for left-handed neutrinos 
while
preserving $SU(2)_L$ gauge symmetry, and the Left-right symmetric model,
which
\cite{LR} requires an $SU(2)_R$ Higgs triplet for 
symmetry breaking with the corresponding left-handed
triplet Higgs field added for the case of explicit $L \leftrightarrow R$
symmetry.

In this paper we study signals for doubly charged Higgs bosons
arising from an $SU(2)_L$ triplet in the 
process $e^-\gamma\to e^+ \mu^-\mu^-$.  
For more details, see reference \cite{us}.
We assume the photon is produced by backscattering a laser from the
$e^+$ beam of an $e^+e^-$ collider \cite{backlaser}.
We consider $e^+e^-$ centre of mass 
energies of $\sqrt{s}=500$, 800, 1000, and 1500~GeV
appropriate to the TESLA/NLC/JLC high energy colliders 
and $\sqrt{s}=3$, 5, and 8~TeV for the CLIC proposal.  In 
all cases we assume an integrated 
luminosity of ${\cal L}=500$~fb$^{-1}$.
Our calculation includes diagrams which would not contribute to on-shell 
production of $\Delta^{--}$'s.  Because the signature of same sign 
muon pairs in the final state is so distinctive and has no SM background, 
we find that the 
process can be sensitive to virtual $\Delta^{--}$'s with masses in
excess of
the centre of mass energy, depending on the strength of the
Yukawa coupling to leptons.  

The $SU(2)_L$ triplet's Yukawa coupling to lepton doublets  is given by 
\begin{equation} 
\label{eq1}
{\cal L}_{Yuk}=-i h_{ll'} \Psi_{l L}^T C \sigma_2 \Delta\Psi_{l' L}+{ h.c.}, 
\end{equation}
where $C$ is the charge conjugation matrix
and $ \Psi_{l L}$ denotes the left-handed lepton
doublet with flavour $l$. Indirect
constraints on $\Delta$ masses and couplings have been obtained from lepton
number 
violating processes
\cite{swartz_and_other}. Rare  decay measurements  
\cite{mu3e_and_other}
yield  very 
stringent restrictions on the
non-diagonal couplings $h_{e\mu}$;
consequently, we choose to neglect all non-diagonal couplings here.
Stringent limits on flavour diagonal couplings 
come from the muonium anti-muonium conversion
measurement \cite{muoniumex} which 
requires that the ratio of the Yukawa coupling, $h$, and Higgs mass, 
$M_\Delta$, satisfy
$h/M_\Delta < 0.44$~TeV$^{-1}$ at 90\% C.L..
These bounds allow the existence of low-mass doubly charged Higgs with 
a small coupling constant.  

Direct search strategies for the $\Delta^{--}$ have been explored for 
hadron colliders \cite{datta_and_others}, with the mass reach at the LHC 
extending to 
$\sim 850$~GeV.
Signatures have also been explored for various configurations of
lepton colliders, including $e\gamma$ colliders. See Reference 
\cite{us} for further references.
The recent calculation of Gregores {\it et al} \cite{gregores}, 
most closely resembles the approach presented here but is restricted 
to resonance $\Delta$ production.

\section{Calculations and Results}
%\label{}

In the process $e^-\gamma \to e^+ 
\mu^-\mu^-$, the signal of 
like-sign muons is distinct and SM background free, 
offering excellent potential for doubly charged Higgs discovery. 
The process proceeds via the production of a positron along 
with a $\Delta^{--}$, with the subsequent $\Delta$ decay into two muons 
as well as through    
additional
non-resonant 
contributions. 
These play an important role in
the reach that one can obtain for doubly charged Higgs masses.

The cross section is a convolution of the backscattered 
laser photon spectrum, $f_{\gamma/e}(x)$ \cite{backlaser}, 
with the subprocess 
cross section, $\hat{\sigma}(e^- \gamma \to e^+ \mu^-\mu^-)$.
Because we are including contributions to the final state that proceed 
via off-shell $\Delta^{--}$'s we must include the doubly-charged Higgs 
boson width.  The $\Delta$ width, however, 
is dependent on the parameters of the model, 
which determine the size and relative importance of various
decay modes.   
To account for the possible 
variation in width without restricting ourselves to 
specific scenarios we calculated the width using
$\Gamma (\Delta^{--}) = \Gamma_b + \Gamma_f$
where $\Gamma_b$ is the partial width to final state bosons and 
$\Gamma_f$ is the partial width into final state fermions.  We consider 
two scenarios for the bosonic width: a narrow width scenario with 
$\Gamma_b=1.5$~GeV and a broad width scenario with $\Gamma_b=10$~GeV. 
These choices represent a reasonable range for various values of the 
masses of the different Higgs bosons.
The partial width to final state fermions is given by
$\Gamma (\Delta^{--}\to \ell^- \ell^-) = \frac{1}{8\pi} 
h^2_{ \ell \ell} M_\Delta$.
Since we assume $h_{ ee} =h_{\mu\mu} =h_{\tau\tau} \equiv h$,
we have 
$\Gamma_f = 3 \times \Gamma (\Delta^{--}\to \ell^- \ell^-) $. Many
studies assume the $\Delta$ decay is entirely into leptons; for small
values of the Yukawa coupling and relatively low $M_{\Delta}$ this leads
to a width which is considerably more narrow than our assumptions for 
the partial width into bosons. Hence, we will also note some results
for 
the case $\Gamma = \Gamma_f$.

We consider two possibilities for the $\Delta^{--}$ signal.  We assume
that either
 all three final state particles are
observed and identified or that  
the
positron is not observed, having been lost down the beam pipe.  
To take into account detector acceptance we restrict 
the angles of the observed particles relative to the beam, 
$\theta_{\mu},\; \theta_{e^+}$, to the ranges $|\cos \theta| \leq 0.9$.
We restrict the particle energies 
$E_{\mu}$, $E_{e^+} \geq 10$~GeV and assumed an identification 
efficiency for each of the
detected final state particles of $\epsilon = 0.9$.

Given 
that the signal for doubly charged Higgs bosons is so distinctive and 
SM background free, discovery would be signalled by even one event.
Because the value of the cross section for the process we consider is
rather sensitive to the $\Delta$ width, the potential for discovery 
of the $\Delta$ is likewise sensitive to this model dependent 
parameter. 
Varying $\Gamma_b$, we find that,
relative 
to $\Gamma_b = 10$~GeV, the case of zero bosonic width has a sensitivity 
to the Yukawa coupling $h$ which is greater by a factor of about 5
\cite{us}.

In Fig. 1
we show 95\% probability (3 event) contours in the $h-M_{\Delta}$ 
parameter space. In each case, we assume the narrow width 
$\Gamma=1.5+\Gamma_f$~GeV case.  
Figure 1a corresponds to the center 
of mass energies  $\sqrt{s}=500$, 
800, 1000, and 1500~GeV,  
for the case of three observed particles in the
final state, whereas Fig. 1b shows the case where only the two muons are
observed.
Figs. 1c and 1d correspond to the energies being 
considered for the CLIC $e^+ e^-$ collider, namely, 
$\sqrt{s}=3$, 5, and 8~TeV,  for the three body and two body
final states, respectively.
In each case, for $\sqrt{s}$ above the $\Delta$ production threshold, 
the process is sensitive to the existence of the $\Delta^{--}$ with 
relatively small Yukawa couplings.  However, when the  $M_\Delta$ 
becomes too massive to be produced the values of the Yukawa couplings 
which would allow discovery grow larger slowly.

\begin{figure}[t]
\centerline{
\begin{minipage}[t]{6.0cm}
%\begin{figure}[htbp]
\vspace*{13pt}
\centerline{
             \hspace{-0.6cm}
\includegraphics[width=5.8cm, height=6cm,angle=-90]{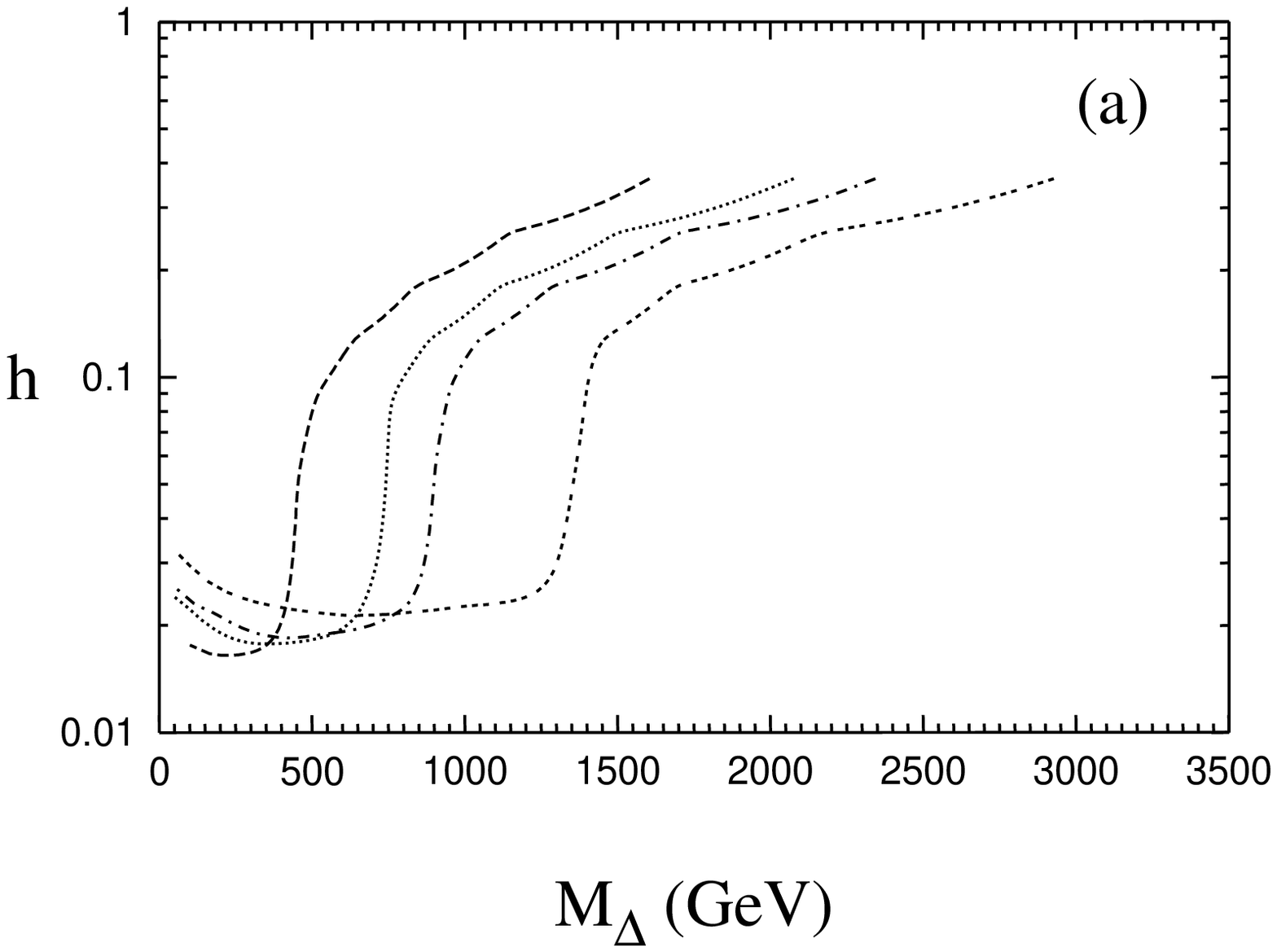}}%,width=6.1cm,clip}}
%\centerline{(6a)}
\vspace*{13pt}
\end{minipage} 
\hspace*{0.5cm}
\begin{minipage}[t]{6.0cm}
\vspace*{13pt}
\includegraphics[width=5.8cm, height=6.0cm,angle=-90]{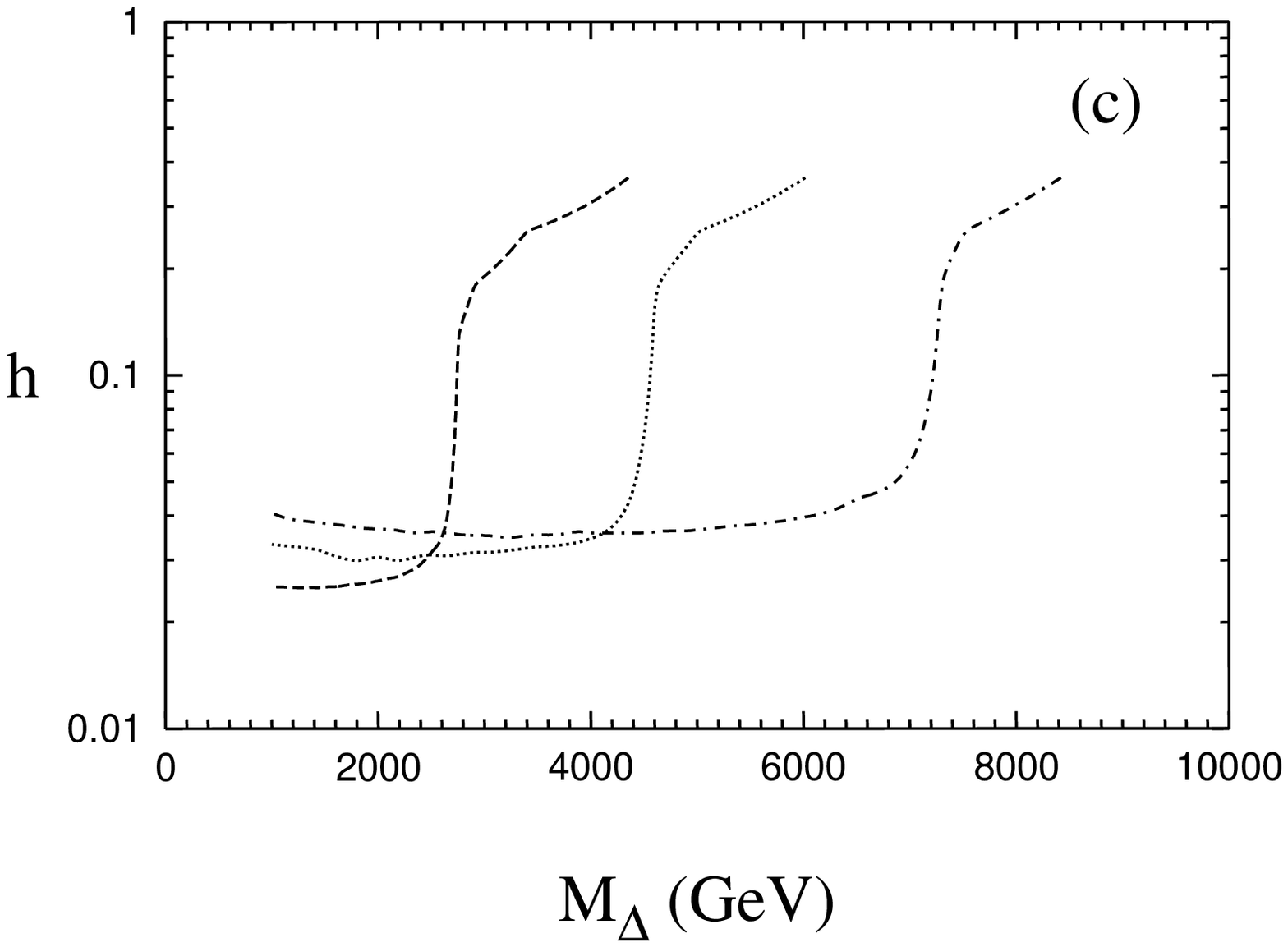}
\vspace*{13pt}
\end{minipage}
}   
\centerline{\hspace*{-0.6cm}
%\newline
\hspace*{0.5cm}
\begin{minipage}[t]{6.0cm}
\vspace*{13pt}
\includegraphics[width=5.8cm, height=6.0cm,angle=-90]{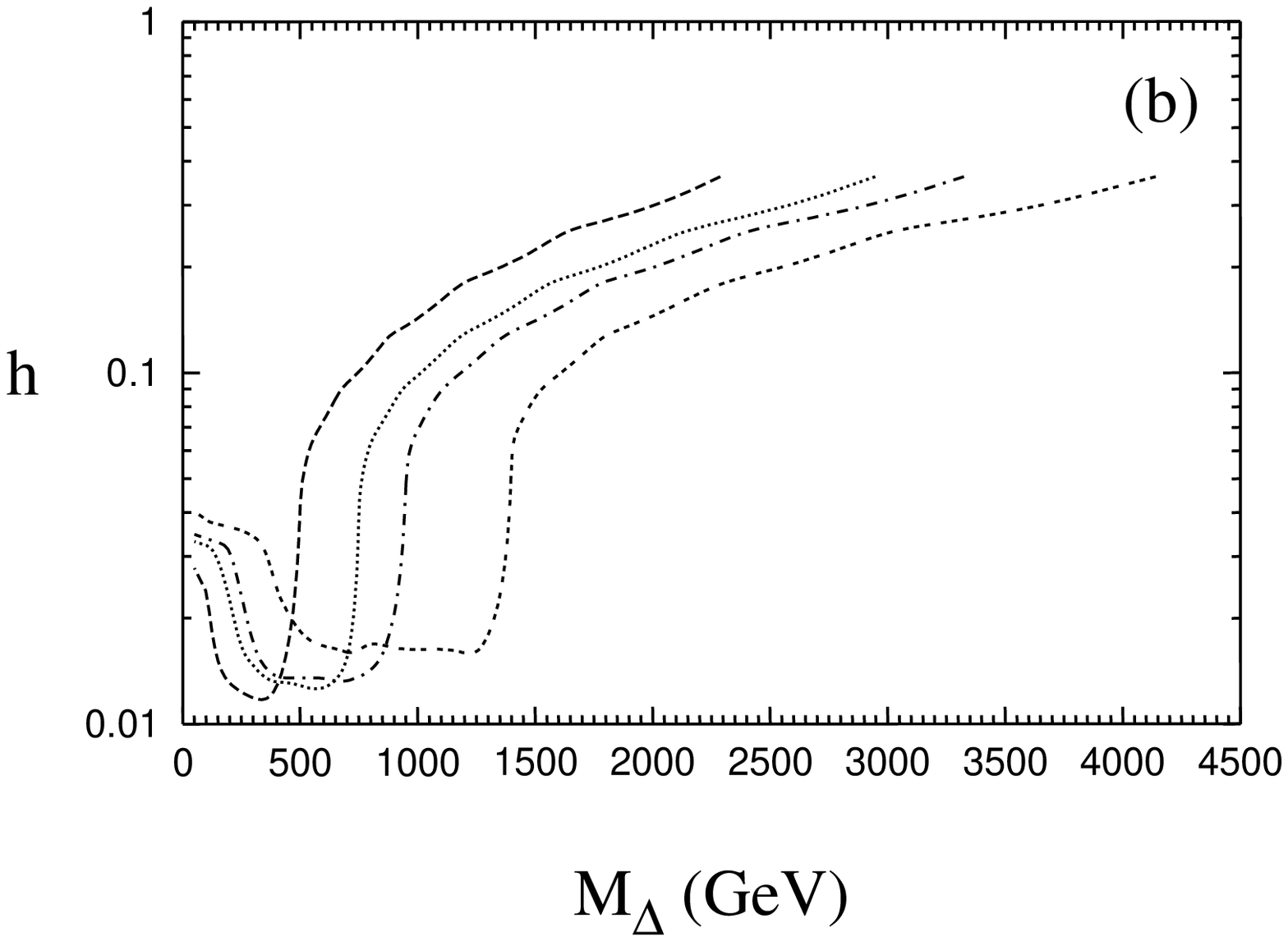}
%\centerline{(6b)}
\vspace*{13pt}
\end{minipage} 
\hspace*{0.5cm}
\begin{minipage}[t]{6.0cm}
\vspace*{13pt}
\includegraphics[width=5.8cm, height=6.0cm,angle=-90]{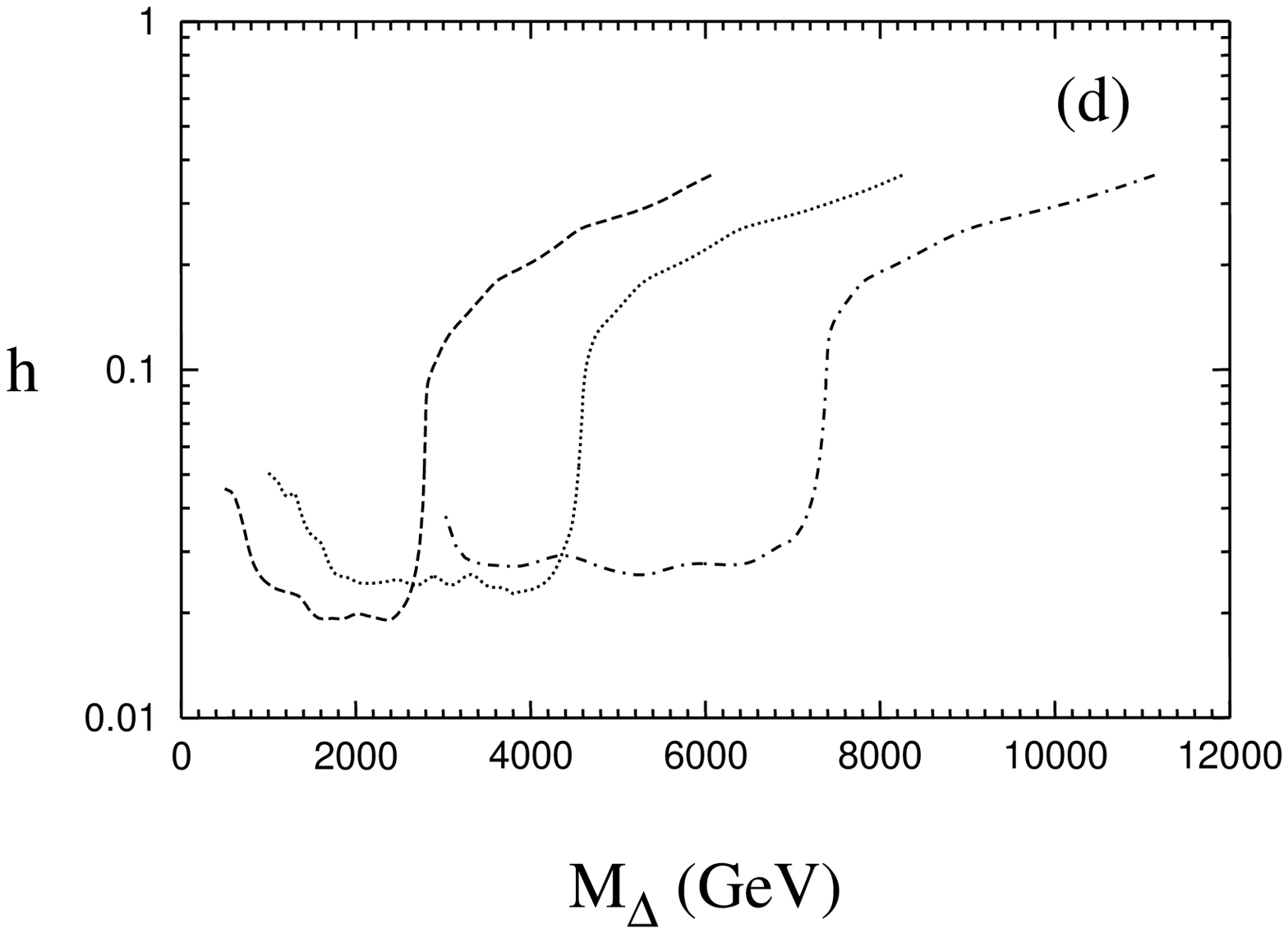}
%\centerline{(7b)}
\vspace*{13pt}
\end{minipage}
         }
\caption{  Discovery limits for the charged Higgs bosons
as a function of Yukawa coupling and $M_{\Delta}$.}
\end{figure}

\section{Summary}
%\label{}

The observation of doubly charged Higgs bosons would represent physics
beyond the SM and, as such, searches for this type of particle
should be part of the 
experimental program of any new high energy facility.  In this paper 
we studied the sensitivity of $e\gamma$ collisions to doubly charged 
Higgs bosons. We found that for $\sqrt{s_{e\gamma}}> M_\Delta$ doubly 
charged Higgs bosons could be discovered for even relatively small 
values of the Yukawa couplings; $h > 0.01$. For larger values of the 
Yukawa coupling the $\Delta$ should be produced in sufficient quantity 
to study its properties.   For values of $M_\Delta$ greater than the 
production threshold, discovery is still possible due to the 
distinctive, background free final 
state in the process  $e\gamma \to e^+ \mu^-\mu^-$ which can proceed 
via virtual contributions from intermediate $\Delta$'s.  Thus, even an 
$e^+e^-$ linear collider with modest energy has the potential to 
extend $\Delta$ search limits significantly higher than can be 
achieved at the LHC.

%\subsection{}
%\subsubsection{}

% figures should be put into the text as floats.
% Use the graphicx package (distributed with LaTeX2e).
% See the LaTeX Graphics Companion by Michel Goosens, Sebastian Rahtz,
% \begin{figure}
% \caption{}
% \label{}
% \end{figure}

% tables follow here or maybe be put in the text
% \begin{table}
% \caption{}
% \label{}
% \end{tabular}
% \end{table}

\begin{acknowledgments}
Work supported by the National Sciences and 
Engineering Research Council of Canada, N.R.
is partially supported by RFFI 01-02-17152.
The authors acknowledgement useful conversations with Dean
Karlen and Richard
Hemingway.
\end{acknowledgments}

\bibliography{example}

\end{document}